\documentclass[aps,twocolumn,showpacs]{revtex4}
\usepackage{graphicx}   
\usepackage{latexsym}   

\newcommand{\beq}{\begin{equation}}
\newcommand{\eeq}{\end  {equation}}

\newcommand{\beqar}{\begin{eqnarray}}
\newcommand{\eeqar}{\end  {eqnarray}}

\newcommand{\SKIP}[1]{}
\newcommand{\bold}[1]{\mbox{\boldmath $#1$}}    

\newcommand{\eg}{{\em e.g.}}			
\newcommand{\etal}{{\em et al.}}		
\newcommand{\ie}{{\em i.e.}}			
\newcommand{\del}{\partial}                     
\newcommand{\GeV}{{\rm GeV}}			
\newcommand{\MeV}{{\rm MeV}}                    
\newcommand{\eps}{\varepsilon}
\newcommand{\bfp}{\bold{p}}       		
\newcommand{\rme}{{\rm e}}                      
\newcommand{\half}{\mbox{${1\over2}$}}          
\newcommand{\third}{\mbox{${1\over3}$}}         
\newcommand{\twothird}{\mbox{${2\over3}$}}	

\newcommand{\code}{{\tt FREYA}}

\begin{document}



\title{Calculation of fission observables through event-by-event simulation}

\author{J{\o}rgen Randrup$^a$ and Ramona Vogt$^{b,c}$}

\affiliation{$^a$Nuclear Science Division, 
Lawrence Berkeley National Laboratory, Berkeley, California 94720, USA}

\affiliation{$^b$Physics Division, 
Lawrence Livermore National Laboratory, Livermore, California 94551, USA}

\affiliation{$^c$Physics Department,
University of California at Davis, Davis, California 95616, USA}

\date{\today}

\begin{abstract}
The increased interest in more exclusive fission observables
has demanded more detailed models.
We present here a new computational model, \code, 
that aims to meet this need
by producing large samples of complete fission events
from which any observable of interest can then be extracted consistently,
including arbitrary correlations.
The various model assumptions are described and 
the potential utility of the model is illustrated
by means of several novel correlation observables.
\end{abstract}


\maketitle

\section{Introduction}

Nuclear fission presents an interesting and challenging physics problem
which is still, about seventy years after its discovery,
relatively poorly understood.
Although much of the key physics involved is understood qualitatively,
a quantitative description is still not in sight,
despite vigorous efforts by many researchers.

Because of its inherent complexity,
fission provides an important testing ground for
both static and dynamical nuclear theories.
Furthermore, fission is also important to society at large
because of its many practical applications,
including energy production and counterproliferation,
topics of current urgency.

Whereas the more traditional treatments of fission 
(see Ref.~\cite{Madland-Nix} and references therein) have sought to describe
only fairly integral fission properties, such as the average energy release
and the average differential neutron yield,
many modern applications require more exclusive quantities,
such as fluctuations in certain observables (\eg\ the neutron multiplicity)
and correlations between between different observables
(\eg\ neutrons and photons).
There is thus a need for developing models that
include the treatment of fluctuations and correlations.

A potentially powerful approach towards meeting this challenge is to develop 
simulation models that can generate samples of complete fission events, 
since a subsequent event-by-event analysis could then provide
any specific correlation observable of interest.
Furthermore, due to the more detailed quantities that they can address,
such models can provide valauble guidance
to experimentalists with regard to which observables are most crucial
for further progress in the understanding of fission.

Relatively recently, Lemaire \etal\ \cite{LemairePRC72} presented 
a Monte-Carlo simulation of the statistical decay of fission fragments
from spontaneous fission of $^{252}{\rm Cf}$ 
and thermal fission of $^{235}{\rm U}$ by sequential neutron emission.
That work demonstrated how fission event simulations, 
in conjunction with experimental data on fission neutrons 
and physics models of fission and neutron emission, 
can be used to predict the neutron spectrum 
and to validate and improve the underlying physics models.

We have developed a conceptually similar calculational framework
within which large samples of complete fission events can be generated,
starting from a fissionable nucleus at a specified excitation energy.
The associated computational code is denoted \code\
({F}ission {R}eaction {E}vent {Y}ield {A}lgorithm).
We present here the model in its most basic form
which, though quite simplistic in many regards,
is already capable of producing interesting results,
as we shall illustrate.
Furthermore, \code\ was employed in a recent study of
sequential neutron emission following neutron-induced fission of $^{240}$Pu
\cite{VPRY}.

In its present early form, \code\ ignores the possibility of neutron emission 
from the nucleus prior to its fission ($n^{\rm th}$ chance fission),
and its applications are therefore limited to lower energies,
such as thermal fission.

In Sects.\ \ref{fission} and \ref{radiation}
we describe above how a single fission event is being simulated
in the pilot version of \code.
By repeating the procedure a large number of times,
we may generate an entire sample of final fission events,
each one consisting of two (slightly excited) residual product nuclei
and the various emitted neutron and photons, 
each one with its associated momentum.
In the development of the numerical code, special care has been taken to
design the various algorithms for fast execution.
As a result, \code\ runs fairly fast, thus making it practical
to generate sufficiently large event samples to permit detailed correlation
analyzes.  
In Sect.\ \ref{examples} we discuss a number of illustrative results.

\section{Fission}
\label{fission}

When the possibility of pre-fission radiation is ignored,
the first physics issues concern how the mass and charge 
of the initial compound nucleus is partitioned among the two fission fragments 
and how the available energy is divided between 
the excitation of the two fragments and their relative kinetic energy.

\subsection{Fission-fragment mass and charge distributions}

In our current understanding of the fission process,
the evolution from the initial compound nucleus to two distinct fission
fragments occurs gradually as a result of a
disipative multidimensional evolution of the nuclear shape.
However, since no quantitatively reliable theory has yet been developed
for this process, we employ empirical evidence 
as a basis for selecting the mass and charge partition.
Thus, the mass and charge partition of the fissioning nucleus $^{A_0}Z_0$
is determined by first selecting the mass partition
from a specified  probability distribution $P(A_f)$
and subsequently selecting the charge partition from 
the associated conditional probabibilty distribution $P_{A_f}(Z_f)$.

In a given event, the mass number $A_f$ of one of the fission fragments
is selected randomly from a probability density $P(A_f)$ 
for which we employ five-gaussian fits to the product mass number distribution 
\cite{YounesPRC64} shifted upwards in mass
to ensure a symmetric distribution of the primary fragments,
\beq
P(A_f)\ =\ \sum_{m=-2}^{m=+2} {\cal N}_m {\cal G}_m(A_f)\ ,
\eeq
where each of the five Gaussians has the form
\beq
{\cal G}_m(A_f)\ =\ \left(2\pi\sigma_m^2\right)^{-{1\over2}}\,
\rme^{-(A_f-\overline{A}_f-D_m)^2/2\sigma_m^2}\ .
\eeq
Contrary to Ref.\ \cite{YounesPRC64}, we are interested in the primary
(\ie\ pre-evaporation) fragment distribution rather than the final
(post-evaporation) product distribution 
and therefore use $\overline{A}_f=\half A_0$.
The fitted values of the normalizations ${\cal N}_m={\cal N}_{-m}$;
the displacements $D_m=D_{-m}$; and
the dispersions $\sigma_m=\sigma_{-m}$ 
have some dependence on the excitation energy $E_0^*$.
Since $\sum_A P(A)=1$ we have ${\cal N}_0+2{\cal N}_1+2{\cal N}_2=1$.

It should be noted that the normalizations ${\cal N}_m$
are not quite correct since
the sum over the integer fragment mass numbers $A_f$ 
does not yield the exact integral of the Gaussian and
the range of the fragment mass numbers $A_f$ is finite.
Because neither of these inaccuracies plays a noticeable role,
we shall ignore them in the present preliminary treatment.
We also note that merely back-shifting the average $\overline{A}_f$
but not reducing the widths $\sigma_m$ 
(to take account of the smearing due to the neutron evaporation)
will lead to a product mass distribution that is a bit too wide 
(since the smearing effect of the neutron evaporation
will, in effect, be taken into account twice).
However, this effect is rather small and is ignored in the present treatment.

For the subsequent selection of the fragment charge number $Z_f$,
we follow Ref.\ \cite{LemairePRC72} and employ a normal distribution,
\beq
P_{A_f}(Z_f)\ \propto\ \rme^{-(Z_f-\overline{Z}_f)/2\sigma_Z^2}\ ,
\eeq
with the condition that $|Z_f-\overline{Z}_f|\leq5\sigma_Z$.
The centroid is determined by demand that the fragments
have the same charge-to-masss ratio as the fissioning nucleus, on average,
$\overline{Z}_f=A_f Z_0/A_0$.
We use the values of the dispersion $\sigma_Z$ measured by 
Reisdorf \etal\ \cite{ReisdorfNPA177},
$0.40$ for $^{236}{\rm U}(n,f)$ and 0.50 for $^{239}{\rm Pu}(n,f)$.
[There appears to be an error (presumably typographical) in the expression (2)
for $P(Z)$ in Ref.\ \cite{LemairePRC72}: the pre-exponential factor
should be a square root in order for $P(Z)$ to be normalized to unity.]

\subsection{Scission energetics}

We obtain the fission energetics by assuming that the two fission fragments
lose contact at a certain scission configuration
which we take to be two coaxial spheroidal prefragments
with a specified tip separation $d$.
For the time being, 
we ignore the nuclear proximity attraction between the two prefragments
as well as any possible relative motion at the time of scission.
These two effects, which counteract one another,
are relatively small but should ultimately be considered.

We introduce some degree of distortion of the pre-fragments
relative to their ground-state shapes,
due to their mutual Coulomb repulsion.
This is done primarily in order to ensure that the resulting
fragment excitations (and hence the neutron multiplicities)
roughly resemble those observed.
Thus, generally, the deformation of the fragment at scission, $\eps_{\rm sc}$,
is larger than that of the ground state, $\eps_{\rm gs}$.
The associated distortion energy is calculated by using the
small-deformation approximation \cite{WDM:Book}, $\delta V
=\mbox{$8\over45$}[E_S^0-\half E_C^0](\eps_{\rm sc}^2-\eps_{\rm gs}^2)$,
which suffices at this early stage of the development.
(Here we use the macroscopic expressions for the surface
energy $E_S^0$ and the Coulomb energy $E_C^0$ for the spherical shape,
as described in App.\ \ref{LD}.)
The distortion moves the prefragment centers apart,
for any fixed tip separation $d$,
and thus lowers the mutual Coulomb repulsion $V^C_{ij}$.

It follows that there are two contributions 
to total excitation of each prefragment,
\beq
E_i^*\ =\ \delta V_i + Q_i\ ,
\eeq
namely the distortion energy $\delta V_i$ 
and the statistical excitaiton (heat) $Q_i$.

The Coulomb repulsion between the two deformed prefragments is calculated
by means of the formula derived in Ref.\ \cite{CohenAP19} for two coaxial, 
uniformly charged spheroids,
\beq
V^C_{ij}\ =\ e^2{Z_iZ_j\over c_i+c_j+d}\, F(x_i,x_j)\ .
\eeq
The factor $F$ is unity for two spheres and larger 
if one or both fragments are prolate.
It depends on the dimensionless deformation measures $x_i$
given by $x_i^2=(c_i^2-b_i^2)/R_i^2$,
where $c_i=R_i[1+\third\eps]/[1-\twothird\eps]^{2/3}$ is the major axis
and $b_i=R_i[1-\twothird\eps]/[1+\third\eps]^{1/3}$ is the minor axis,
while $R_i$ is the average radius of the fragment.

Once the fragments have lost contact, they are accelerated
by their mutual Coulomb repulsion and their shapes relax
to their equilibrium forms.
The scission distortion energies are converted into additional
statistical excitations of the respective fragments.
We assume that these processes
have been completed before the de-excitation processes begin.

With the (significant) simplifications described above,
we have the following simple energy relations 
for any particular fission channel, 
$^{A_0}\!Z_0\to\,^{A_L}\!Z_L+\,^{A_H}\!Z_H$,
\beqar\nonumber
M_0^*\ =\ M_0^{\rm gs}+E_0^*
&=& M_L^{\rm gs} +E_L^* +M_H^{\rm gs} +E_H^* +V_{LH}^C\\
&=& M_L^*+M_H^*+K_{LH}\ .
\eeqar
Here $M_i^{\rm gs}$ is the ground-state mass of the nucleus $^{A_i}Z_i$,
$i=0,L,H$, and $E_i^*$ is its excitation, 
so $ M_i^*=M_i^{\rm gs}+E_i^*$ is its total mass.
[The ground-state masses are taken 
from the compilation by Audi {\em et al.}~\cite{Audi},
supplemented by calculated masses by M{\"o}ller {\em et al.}~\cite{MNMS}
where no data are available.]
Furthermore, $V_{LH}^C$ is the Coulomb repulsion 
between the two light and heavy fragments at scission.
This energy is, by {\em fiat}, fully converted into relative kinetic energy 
of the two receding fission fragments, $K_{12}$.
Thus, in addition to ignoring any possible post-scission dissipation,
we also disregard any angular-momentum effects.
While these effect are expected to be small, 
it might be of interest to include them at a later time.
The $Q$-value associated with the particular fission channel is given by
\beq
Q_{0\to LH}\ =\ M_0^{\rm gs}+E_0^* -M_L^{\rm gs} -M_H^{\rm gs}\ 
=\ K_{LH} +E_L^* +E_H^*\ .
\eeq

\subsection{Thermal fluctuations}

Once the scission configuration is known,
its average total internal (statistical) excitation energy, $\overline{Q}$, 
can be readily obtained,
\beq
\overline{Q}\ \equiv\ \overline{Q}_L+\overline{Q}_H\
=\ M_0^{\rm gs}+E_0^* -M_L^{\rm sc}-M_H^{\rm sc}-V_{LH}^C\ ,
\eeq
where $M_i^{\rm sc}=M_i^{\rm gs}+\delta V_i$ is the mass of the
distorted prefragment of the scission configuration.
We assume that this internal energy $\overline{Q}$ 
is partitioned statistically between the two prefragments, 
as would be the case when the two are in mutual thermal equilibrium.
Thus, on the average, the total excitation energy is divided in proportion to 
the respective heat capacities.
These in turn are characterized by the Fermi-gas level-density parameters $a_i$
which are approximately proportional to the fragment masses $A_i$;
we use the values calculated in Ref.~\cite{KouraNPA674} (see App.\ \ref{aA}).
[We note that those calculations were made for nuclei 
in their ground-state shapes,
whereas the scission pre\-fragments are distorted and may thus have different
effective level-density parameters.]
The mean excitation in a nucleus is assumed to be $\overline{Q}_i=a_iT_i^2$.
so the heat capacity is $\del\overline{Q}_i/\del T_i=2a_i T_i\propto a_i$.
Since the two prefragments in the scission configuration
have a common temperature, 
$T_{LH}=[\overline{Q}/(a_L+a_H)]^{1/2}=[\overline{Q}_i/a_i]^{1/2}$,
we use $\overline{Q}_i=a_iT_{LH}^2$.

The fluctuations in the statistical excitation $Q_i$ 
are given by the associated 
thermal variances, $\sigma_i^2=2\overline{Q}_i T_{LH}$.
The fluctuations $\delta Q_i$ are therefore sampled from 
normal distributions with variances $\sigma_i^2$.
The prefragment excitations in a given event are then
$Q_i=\overline{Q}_i+\delta Q_i$.

As a result of the fluctuations in the statistical excitation
energies of the individual prefragments, $Q_i$, 
the combined statistical excitation energy, $Q=Q_L+Q_H$, will also fluctuate.
This fluctuation in turn implies a compensating fluctuation
in the total fragment kinetic energy, 
so that $K_{LH}=\overline{K}_{LH}+\delta K_{LH}$ where
\beq
\overline{K}_{LH} = V_{LH}^C\ ,\,\,\
\delta K_{LH}=-\delta Q_L-\delta Q_H\ .
\eeq

We note that the resulting thermal distribution of heat in each prefragment
is approximately gaussian,
\beq
P_i(Q_i)\ \approx\ (2\pi\sigma_i^2)^{-{1\over2}}
	{\rm e}^{-(Q_i-\overline{Q}_i)^2/2\sigma_i^2}\ .
\eeq
Consequently, the distribution of the combined amount of heat
in both fragments, $Q=Q_L+Q_H$, is also approximately gaussian
and the associated variance is the sum of the individual variances,
$\sigma_Q^2=\sigma_L^2+\sigma_H^2$.
Energy conservation implies that
the distribution of the total kinetic energy $K_{LH}$
is a gaussian with the same width, $\sigma_K=\sigma_Q$,
as was assumed in Ref.\ \cite{LemairePRC72}.

It is physically reasonable that the partioning of the total energy 
between kinetic energy and internal excitation fluctuates
because the evolution of the fissioning system from saddle to scission
is a dissipative process.
The associated conversion of the collective energy to heat
is the result of many elementary stochastic processes.
The fluctuation-dissipation theorem then relates the average energy loss
(the dissipation) to the associated fluctuation.
Energy conservation demands that the fluctuations in the kinetic energy
are exactly the opposite of those in the internal excitation.
These, in turn, are given by the above thermal expressions insofar as
statistical equilibrium is maintained during the shape evolution
from saddle to scission.
[We ignore the possibility that the scission configuration itself 
might also fluctuate from one event to another for a given fission channel.]

Once the relative kinetic energy $K_{LH}$ has been obtained as described above,
the magnitude of the relative momentum, $p_{LH}$,
of the fully accelerated fragments is then determined.
Since the kinetic energy is relatively small
($K_{LH}\approx200\,\MeV$, while $M_0^*>200\,\GeV$), 
we may safely assume that $K_{LH}\ll M_i^*$ 
and use non-relativistic kinematics, $p_{LH}^2=2\mu_{LH}K_{LH}$,
where the reduced fragment mass is $\mu_{LH}=M_L^*M_H^*/(M_L^*+M_H^*)$
with $M_i^*=M_i^{\rm sc}+Q_i=M_i^{\rm gs}+\delta V_i+Q_i$
being the total mass of the excited prefragment.
Ignoring any angular momentum effects,
we select the fission direction $\bold{\hat{V}}$ randomly.
The fragment momenta are then $\bold{P}_L=p_{LH}\bold{\hat{V}}$ and
$\bold{P}_H=-p_{LH}\bold{\hat{V}}$, in the frame of the fissioning nucleus.

\section{Post-fission radiation}
\label{radiation}

As mentioned above, we assume that the two excited fragments 
do not begin to de-excite
until after they have been fully accelerated by their mutual Coulomb replusion
and their shapes have reverted to their equilibrium form,
which we take to be those of their ground states.
[In principle, the equilibrium shape of a nucleus depends on its excitation
since both shell effects and surface tension are temperature dependent, 
but we have ignored this relatively minor complication at this time.]
Furthermore, we ignore the possibility of charged-particle emission
from the fission fragments.

Each of the fully relaxed and accelerated fission fragments
typically emits one or more neutrons 
as well as a (larger) number of photons.
We assume that neutron evaporation has been completed
(\ie\ no further neutron emission is energetically possible)
before photon emission sets in.
This simplifying assumption obviates the need for knowing the
ratio of the widths, $\Gamma_\gamma(E_i^*)/\Gamma_n(E_i^*)$.

\subsection{Statistical evaporation of neutrons}

We treat post-fission neutron radiation 
by iterating a simple treatment of a single neutron evaporation,
until no further neutron emission is energetically possible.

Statistical neutron evaporation is but one example 
of a general two-body decay.
In the present case, the initial body is an excited nucleus
with a total mass equal to its ground-state mass
plus its excitation energy, $M_i^*=M_i^{\rm gs}+E_i^*$.
The $Q$-value for neutron emission is then 
$Q_{\rm n}=M_i^*-M_f^{\rm gs}-m_{\rm n}$,
where $M_f^{\rm gs}$ is the ground-state mass of the daughter nucleus and
$m_{\rm n}$ is the mass of the (unexcitable) ejectile (the neutron).
The $Q$-value equals the maximum possible excitation energy
of the daughter nucleus, which is achieved for vanishing final 
relative kinetic energy, $Q_{\rm n}=E_f^{\rm max}$,
which would be obtained if the emitted neutron had no kinetic energy.
It is related to the associated maximum daughter temperature $T_f^{\rm max}$
by $a_f(T_f^{\rm max})^2=E_f^{\rm max}$,
where $a_f$ is the level density parameter of the daughter nucleus
(see App.~\ref{aA}).

\subsubsection{Spectral profile}

Once the $Q$-value is known,
it is straightforward to sample the kinetic energy 
of an evaporated neutron,
assuming that it is isotropic in the rest frame of the emitting nucleus.
We first note that the kinetic energy of the neutron has the form
$\epsilon_{\rm n}=p_{\rm n}^2/2m_{\rm n}$ 
while $v_{\rm n}\propto\sqrt{\epsilon_{\rm n}}$
(non-relativistically) so that 
$d^3\bfp_{\rm n}\propto\sqrt{\epsilon_{\rm n}}d\epsilon_{\rm n}$ 
for isotropic emission.
The differential distribution is then \cite{Weisskopf,BlattWeisskopf}
\beqar\label{dNdeps}\nonumber
{d^3\nu\over d^3\bfp_{\rm n}}\,d^3\bfp_{\rm n} &\propto& 
	\sqrt{\epsilon_{\rm n}}\,
	{\rm e}^{-\epsilon_{\rm n}/T_f^{\rm max}}
	\sqrt{\epsilon_{\rm n}} d\epsilon_{\rm n}\,d\Omega\\
&=& \epsilon_{\rm n}\,{\rm e}^{-\epsilon_{\rm n}/T_f^{\rm max}}
	d\epsilon_{\rm n}\,d\Omega\ ,
\eeqar
in the rest frame of the emitting nucleus.
The form $\sqrt{\epsilon_{\rm n}}\exp(-\epsilon_{\rm n}/T)$ can be 
understood as the product of the thermal occupancy of the neutron,
$\propto\exp(-\epsilon_{\rm n}/T)$,
and its normal speed $v_{\rm n}\propto\sqrt\epsilon_{\rm n}$
which introduces a bias in favor of those neutrons that are
moving perpendicular to the nuclear surface.

The kinetic energy of the evaporated neutron, $\epsilon_{\rm n}$,
is sampled by means of a specific fast algorithm 
that is described in App.\ \ref{eps}.
We note that the form of the energy spectrum implies
that the evaporated neutron has a mean (relative) kinetic energy of
$\langle\epsilon_{\rm n}\rangle=2T_f^{\rm max}$
and an associated variance of $2(T_f^{\rm max})^2$.
These expressions apply to the particular stage of the evaporation chain.
Generally, the first neutron evaporated from the fragment
will tend to have a higher energy than the second one, and so on.

\subsubsection{Kinematics}

Although relativistic effects are very small, we wish to take them into account
in order to ensure exact conservation of energy and momentum,
which is convenient for code verification purposes.
We therefore take the above sample value $\epsilon$ to represent the
{\em total} kinetic energy in the rest frame of the mother nucleus,
\ie\ it is the kinetic energy of the emitted neutron
{\em plus} the recoil energy of the residual daughter nucleus.
The excitation energy in the daughter nucleus is then given by
\beq
E_f^*\ =\ Q_{\rm n}-\epsilon_{\rm n}\ .
\eeq
Since relativistic mass of the daughter nucleus is $M_f^*=M_f^{\rm gs}+E_f^*$,
it is possible to calculate the momenta of the emitted neutron
and the excited daughter as follows.

Generally, if a particle of mass $M$ decays into two particles of masses
$m_1$ and $m_2$, those two particles are emitted back-to-back 
in the rest frame of the initial particle,
with their momenta having equal magnitudes.
Denoting this common momentum magnitude by $p$, 
application of elementary energy conservation yields
\beq
M\ =\ E_1 + E_2\ =\ [m_1^2+p^2]^{1/2}\ +\ [m_2^2+p^2]^{1/2}\ ,
\eeq
from which the magnitude $p$ can be readily obtained,
\beq
4M^2p^2\ =\ [M^2-(m_1+m_2)^2][M^2-(m_1-m_2)^2]\ .
\eeq
The individual energies, $E_i=[p^2+m_i^2]^{1/2}$,
may then be obtained subsequently.
We employ the above formula with $M=M_i^*$, 
$m_1=m_{\rm n}$ and $m_2=M_f^*=M_f^{\rm gs}+Q_{\rm n}-\epsilon_{\rm n}$.

Assuming that the emission is isotropic
(which follows from the neglect of angular-momentum effects),
we may readily sample the direction of relative motion $(\vartheta,\varphi)$.
The momentum of the ejectile is then
\beq
\bfp_{\rm n}=(	p\,\cos\!\varphi\,\sin\!\vartheta,
		p\,\sin\!\varphi\,\sin\!\vartheta,
		p\,\cos\!\vartheta),
\eeq
while the recoil momentum of the residue is the opposite,
$\bold{P}_f=-\bfp_{\rm n}$.
These momenta are in the two-body CM frame,
the frame of the mother nucleus, which would generally be moving.
We therefore need to boost these momenta to the overall reference frame
(see App.\ \ref{boost}).

The emission procedure described above may be repeated until 
no further neutron emission is energetically possible.
That happens when $E_f^*<S_{\rm n}$,
where $S_{\rm n}$ is the neutron separation energy for the daughter nucleus,
$S_{\rm n}=M(^AZ)-M(^{A-1}Z)-m_{\rm n}$.

\subsection{Statistical emission of photons}

Although, at this initial stage, our main focus is on neutron evaporation,
we wish to also include an approximate treatment of photon emission.
For this purpose we disregard nuclear structure effects
and treat the post-evaporation photon cascade 
in a manner that is similar to the neutron emission described above.
Clearly, this part can be refined by taking account of the specific 
level structure in the fission fragments.
Because the photon is massless, we introduce an energy cutoff (see below).

Furthermore, the vanishing photon mass causes it to be ultrarelativistic
with $p_\gamma c=\epsilon_\gamma$ and $v_\gamma=c$.  Consequently,
\beq\label{dGdeps}
{d^3N_\gamma\over d^3\bfp_\gamma}\,d^3\bfp_\gamma\ \propto\ \epsilon_\gamma^2
{\rm e}^{-\epsilon_\gamma/T_f} d\epsilon_\gamma\,d\Omega\ ,
\eeq
as was also used in Ref.~\cite{LemairePRC73}.
For the first photon to be emitted,
$T_f$ is the temperature in the nucleus right after the 
last neutron was evaporated, $a_fT_f^2=E_f^*$,
and generally it is the temperature 
{\em before} the next photon is emitted.

The photon energy $\epsilon_\gamma$ is sampled by a fast algorithm 
(see App.~\ref{eps}) 
and the nuclear excitation energy is reduced correspondingly,
$(E_f^*)'=E_f^*-\epsilon_{\rm n}$.
The spectral shape (\ref{dGdeps}) yields an average photon energy 
of $\langle\epsilon_\gamma\rangle=3T_f$ and an associated variance of $3T_f^2$,
for a fixed value of $T_f$.
Since, in principle, the continuous form of the spectrum
leads to an infinite number of ever softer photons,
we keep track of only those with an energy above a specified threshold,
$\epsilon_\gamma^{\rm min}=200\,{\rm keV}$.
For photons above that threshold,
the emission direction is sampled uniformly over $4\pi$
and a Lorentz boost is performed to express the emitted photon
and the nuclear residue in the overall reference frame.

This procedure is iterated until the nuclear excitation energy
falls below the specified minimum value $\epsilon_\gamma^{\rm min}$.

\section{Illustrative results}
\label{examples}

Here, we wish to illustrate the utility of single-event models like \code\
by presenting a number of correlation observables 
that could not be addressed with earlier codes
which have tended to focus on more inclusive quantities.
Obviously, the present preliminary version of \code\ 
involves a number of simplifying approximations and, consequently,
the results cannot be expected to be numerically accurate.
Certainly, for the most common observables,
such as average multiplicities and spectra,
the most reliable results can undoubtedly be obtained
from the well-tuned codes that have long been available.
We expect that event simulation codes will, in due course,
achieve a similar level of accuracy.
Meanwhile, they may serve as useful supplements
with which is will be possible to address more detailed observables
on an approximate level.

While the main purpose here is to illustrate the kind of novel information 
that can be accessed with \code, we wish to first show a number of more
familiar observables.
Throughout we consider fission induced by thermal neutrons
on $^{235}$U and $^{239}$Pu.
Fission induced by higher-energy neutrons is not considered,
since the possibility of pre-fission neutron emission
(and the associated $n^{\rm th}$ chance fission)
has not yet been included.

\begin{figure}          
\includegraphics[angle=0,width=3.1in]{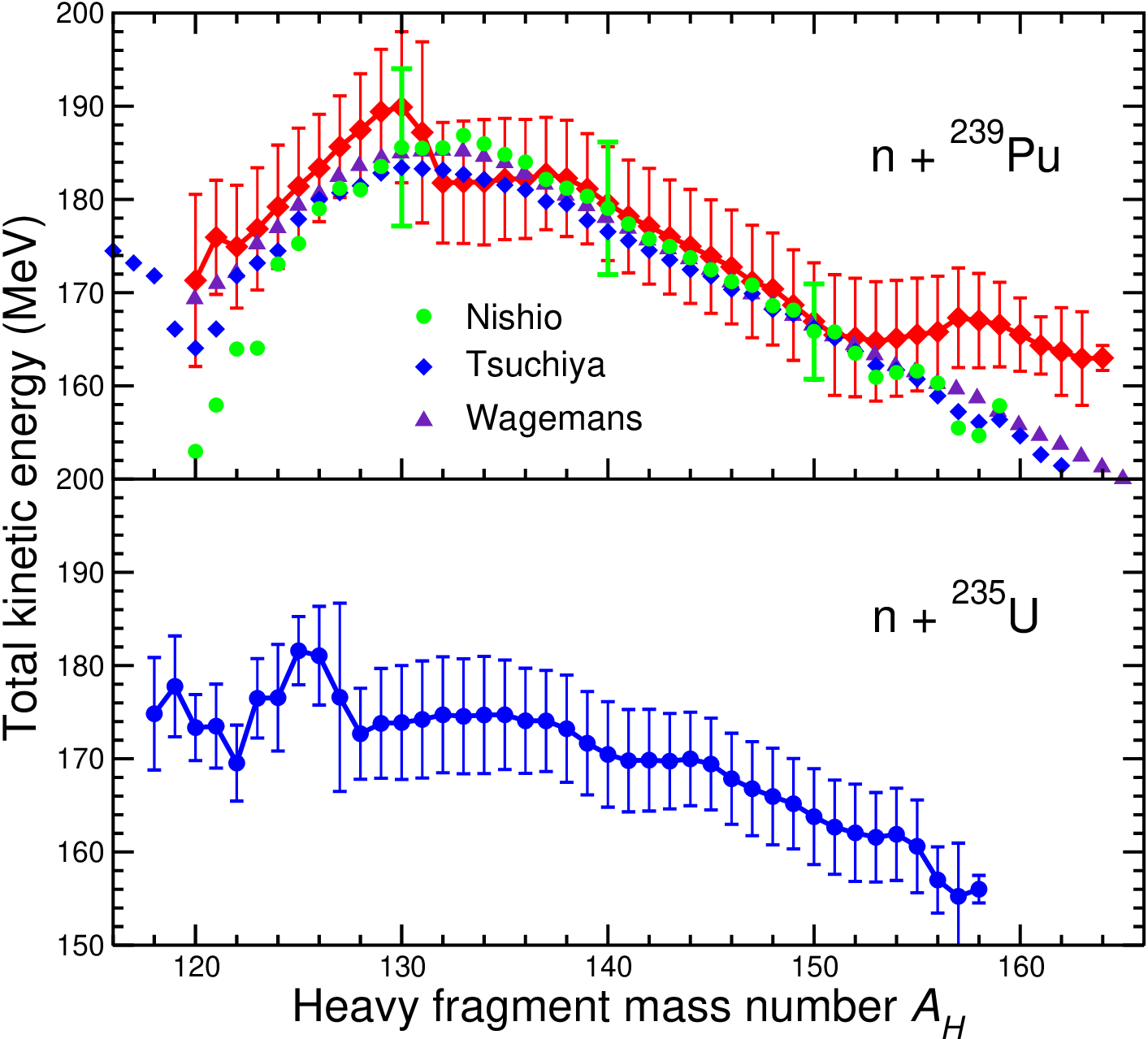}
\caption{
The mean total kinetic energy, $\overline{K}_{\rm tot}$,
of the two fission fragments, and the associated dispersion (bars),
as a function of the mass number of the heavy fragment,
$A_H$, for $0.53\,\MeV\,{\rm n}$ on $^{235}$U ({bottom}) 
and $^{239}$Pu ({top}).
The data from Nishio \cite{Nishio} (with a few representative dispersions), 
Tsuchiya \cite{TsuchiyaPu}, and Wagemans \cite{WagemansPu} are shown.
The dispersions reflect the width of the kinetic energy distribution
and are not (experimental or theoretical) uncertainties.
[Color online.]}\label{f:KtotAH}
\end{figure}            

\begin{figure}          
\includegraphics[angle=0,width=3.1in]{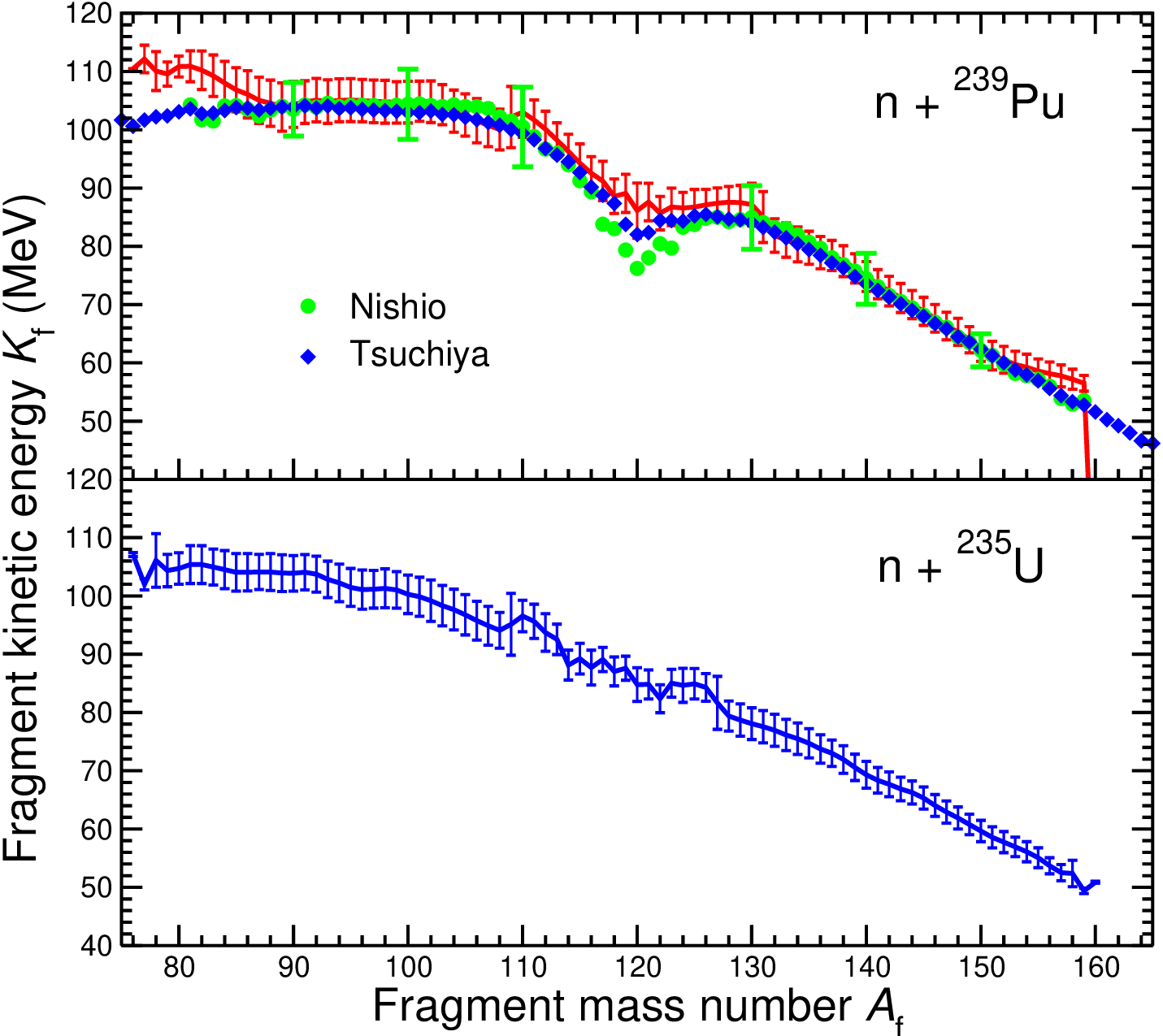}	
\caption{
The mean kinetic energy of a single fragment
and the associated dispersion (bars)
as a function of its mass number $A_f$,
for $0.53\,\MeV\,{\rm n}$ on $^{235}$U ({bottom}) 
and $^{239}$Pu ({top}).
Pu data from Nishio \cite{Nishio} and Tsuchiya \cite{TsuchiyaPu} are shown.
[Color online.]}\label{f:KAf}
\end{figure}            

\subsection{Fission fragments}

The most basic observable is perhaps the product mass distribution $P(A_p)$
which, by design, matches the fits to the observed data and thus need not be
displayed.

We therefore start by considering the kinetic energies 
of the fission fragments.
Figure \ref{f:KtotAH} shows the combined kinetic energy of both fragments,
$K_{\rm tot}$, as a function of the mass number of the heavy fragment, $A_H$,
while Fig.\ \ref{f:KAf} shows the kinetic energy of a single fragment
as a function of its mass number $A_f$.
These results exhibit the general observed features,
though the detailed behavior is not yet expected to be accurate.

The figures show the mean values of the kinetic energies 
as well as the associated dispersions.
A quick comparison of the two figures suggests that the variance of the
total kinetic energy is generally larger than the sum of the variances
of the individual kinetic energies.
This striking feature is an elementary consequence of momentum conservation.
Since the two fragments emerge with opposite momenta,
the fluctuations in their kinetic energies are closely correlated.
As a result, the sum of the variances of the two individual fragment energies,
$K_L$ and $K_H$, is significantly smaller than the variance in the combined
fragment energy $K_{LH}=K_L+K_H$, namely
$\sigma^2(K_L)+\sigma^2(K_H)=[1-2A_LA_H/A_0^2]\,\sigma^2(K_{LH})$.
In particular, for a symmetric split, $A_L\!=\!A_H$, we have
$\sigma(K_i)=\half\sigma(K_{LH})$
hence $\sigma(K_{LH})^2=2[(\sigma(K_L)^2+\sigma(K_H)^2]$.

\begin{figure}          
	\includegraphics[angle=0,width=3.1in]{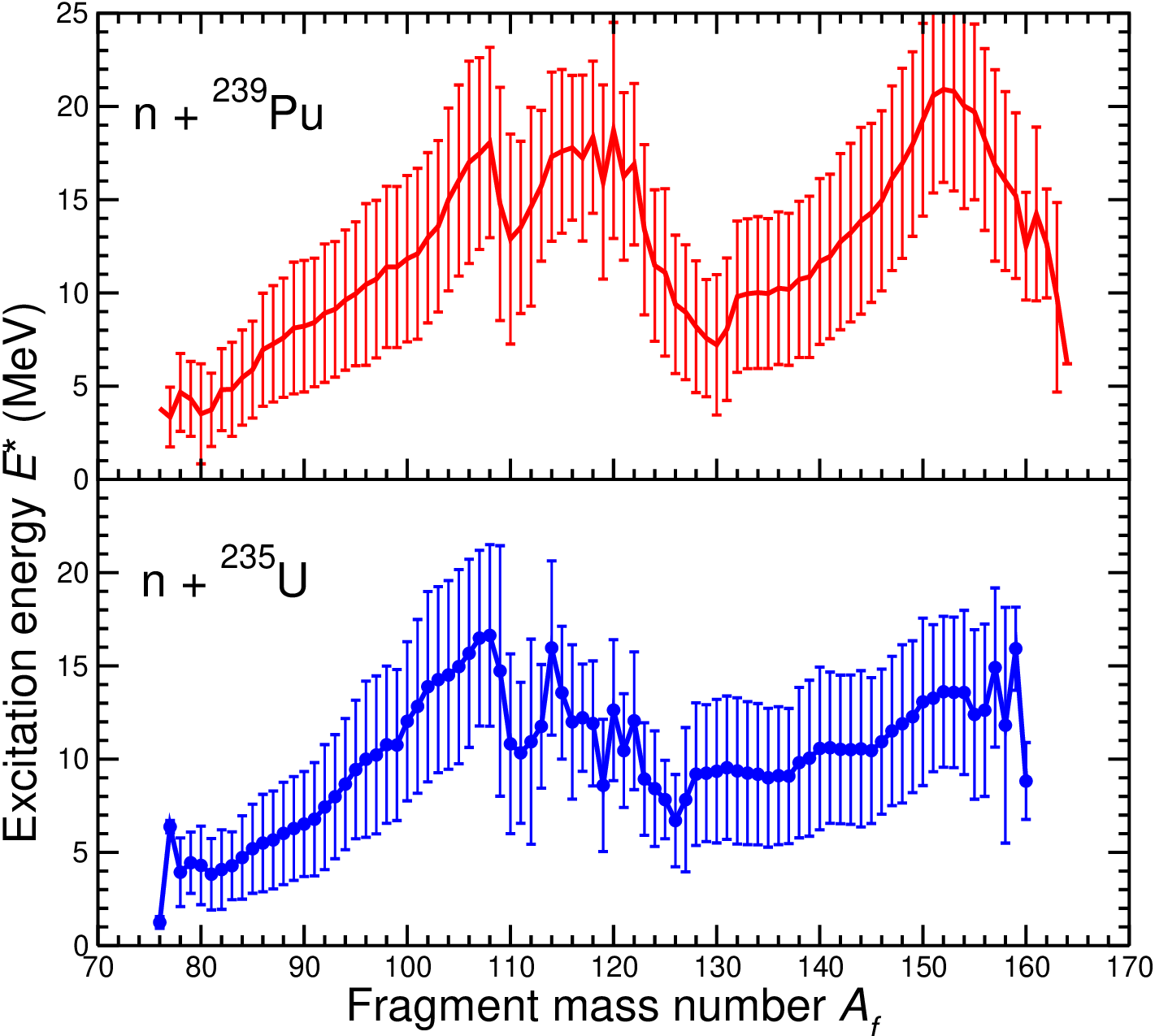}
\caption{
The mean excitation energy $\overline{E}^*$ (curves) of a fission fragment
and the associated dispersion (bars), as a function of the mass number $A_f$
for $0.53\,\MeV\,{\rm n}$ on $^{235}$U ({bottom}) and $^{239}$Pu ({top}).
[Color online.]}\label{f:EAf}
\end{figure}

While the total excitation of the emerging fragments is related to their
total kinetic energy by energy conservation,
its partition is less straightforward,
depending both on the relative heat capacities (\ie\ level densities)
and the scission fluctuations.
Figure \ref{f:EAf} shows the mean fragment excitation $\overline{E}^*_f$
together with the associated dispersion, 
as a function of the fragment mass number $A_f$.
In the present model, the division of the available energy
between kinetic and excitation is sensitive to 
the degree of distortion of the scission pre-fragments,
a property that in turn depends on 
the shell structure of the specific nuclides.

\subsection{Neutron multiplicities}

\begin{figure}          
\includegraphics[angle=0,width=3.1in]{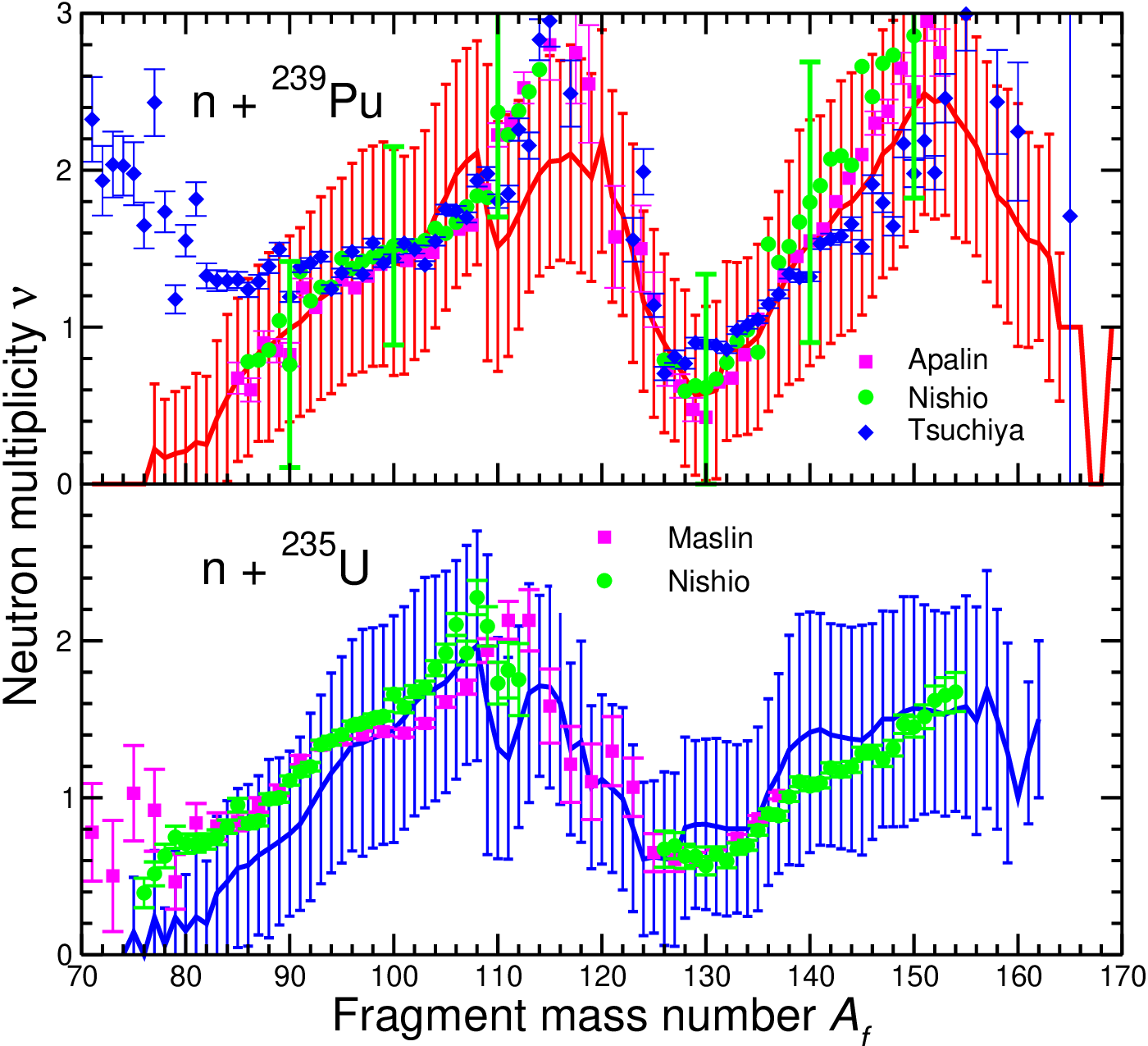}	
\caption{
The mean number of neutrons (curve) 
resulting from fission fragment of a particular mass number $A_f$, 
with the associated dispersion indicated (bars),
for $0.53\,\MeV\,{\rm n}$ on $^{235}$U ({bottom}) and $^{239}$Pu ({top}).
Data from Maslin \cite{Maslin}, Nishio \cite{Nishio},
 and Tsuchiya \cite{TsuchiyaPu} are shown.
[Color online.]}\label{f:mAf}
\end{figure}

The fission fragment excitation energies $E^*(A_f)$ (see Fig.\ \ref{f:EAf}) 
largely determine the multiplicities of evaporated neutrons $\nu(A_f)$.
This correspondance is clearly seen in Fig.\ \ref{f:mAf}
which shows the mean neutron multiplicity $\overline{\nu}(A_f)$
and the associated dispersion $\sigma_\nu(A_f)$.
We note that the observed sawtooth shape is roughly reproduced,
though the detailed behavior is not completely satisfactory.

\begin{table}
\begin{tabular}{l|cccc}
\hline
~ & ~~~$~~~\overline{\nu}_L$~~~ & ~~~$\overline{\nu}_H $~~~ 
& ~~~$\overline{\nu}$~~~& ~$C_{LH}$~\\
\hline
n+$^{239}$Pu & 1.53 & 1.43 & 2.96 & -0.19\\
n+$^{235}$U  & 1.23 & 1.23 & 2.47 & -0.12\\
\hline
\end{tabular}
\caption{The mean number of neutrons emitted from either the light fragment,
$\overline{\nu}_L$, the heavy fragment, $\overline{\nu}_H$, 
or either fragment, $\overline{\nu}=\overline{\nu}_L+\overline{\nu}_H$,
in fission events induced by thermal neutrons on $^{235}$U and $^{239}$Pu.
Also shown is the correlation coefficient 
$C_{LH}\equiv\sigma(\nu_L,\nu_H)/[\sigma(\nu_L)\sigma(\nu_H)]$.
}\label{t:nu}
\end{table}

\begin{figure}          
\includegraphics[angle=0,width=3.1in]{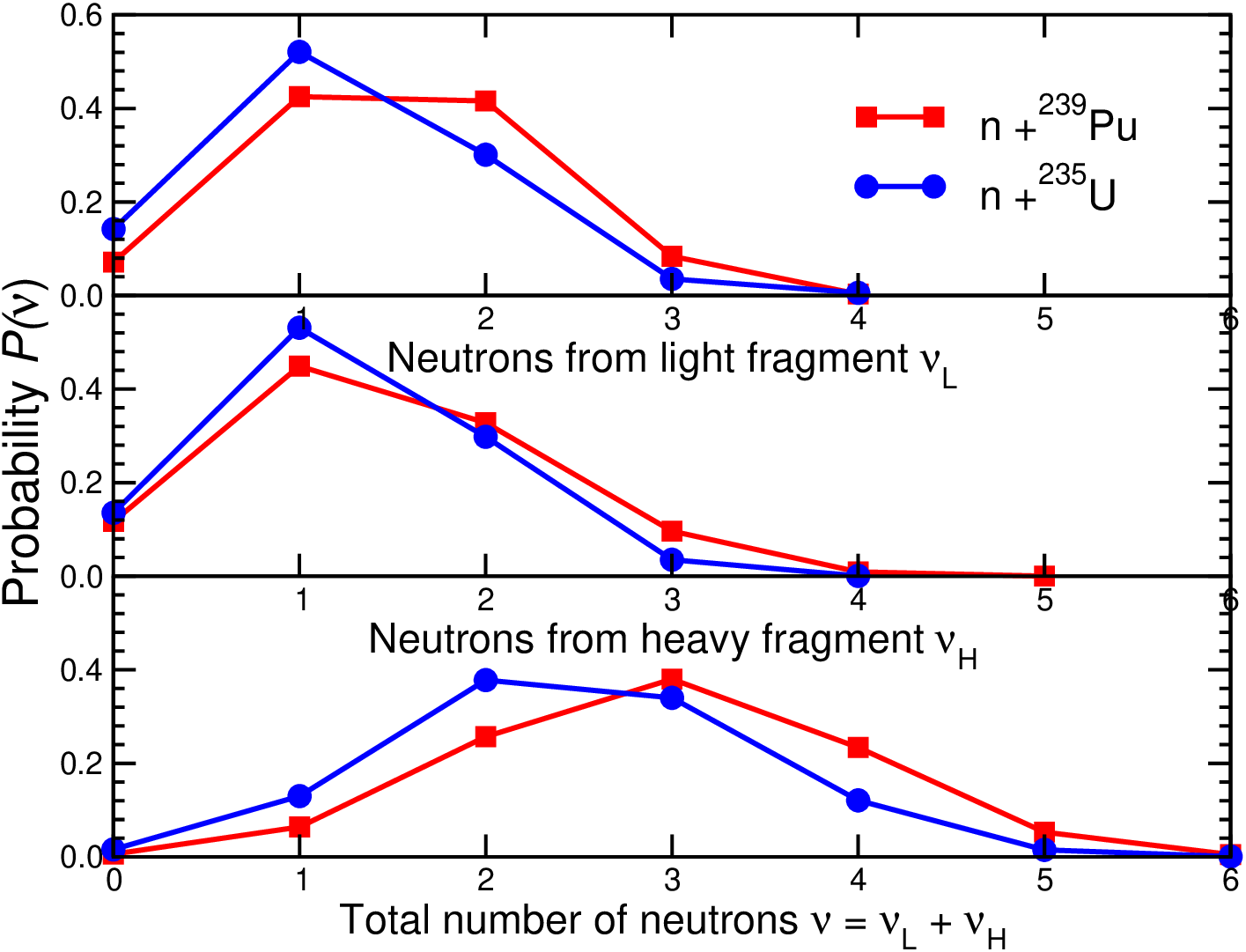}
\caption{
Multiplicity distributions for neutrons emitted by the light ({top}),
the heavy ({\em middle}), or either ({bottom}) fragment resulting from
thermal-neutron induced fission of
$^{235}$U ({\em circles, blue}) and $^{239}$Pu ({\em squares, red}).
[Color online.]}\label{f:nDist}
\end{figure}

The overall neutron multiplicity distribution $P(\nu)$
is shown in Fig.\ \ref{f:nDist}.
This figure also shows the separate multiplicity distributions 
$P(\nu_L)$ and $P(\nu_H)$ for the number neutrons $\nu_L$ and $\nu_H$
that were emitted by the light or the heavy fragment, respectively,
a quantity that is difficult to obtain experimentally.
The associated average multiplicities are shown in Table \ref{t:nu}
($\overline{\nu}_L\equiv\langle\nu_L\rangle$, {\em etc.}),

We note that the light fragment tends to emit more than its ``fair share''
of neutrons, a reflection of the fact that 
the excitation energy is not divided solely in proportion to mass.
Futhermore, as the correlation coefficient $C_{LH}$ shows,
there is a slight anticorrelation between $\nu_L$ and $\nu_H$.
This feature is presumably a result of the anticorrelation 
between the excitations of the two partner fragments
caused by the thermal fluctuations of the heat partition at scission.

\begin{figure}[t]          
\includegraphics[angle=0,width=3.1in]{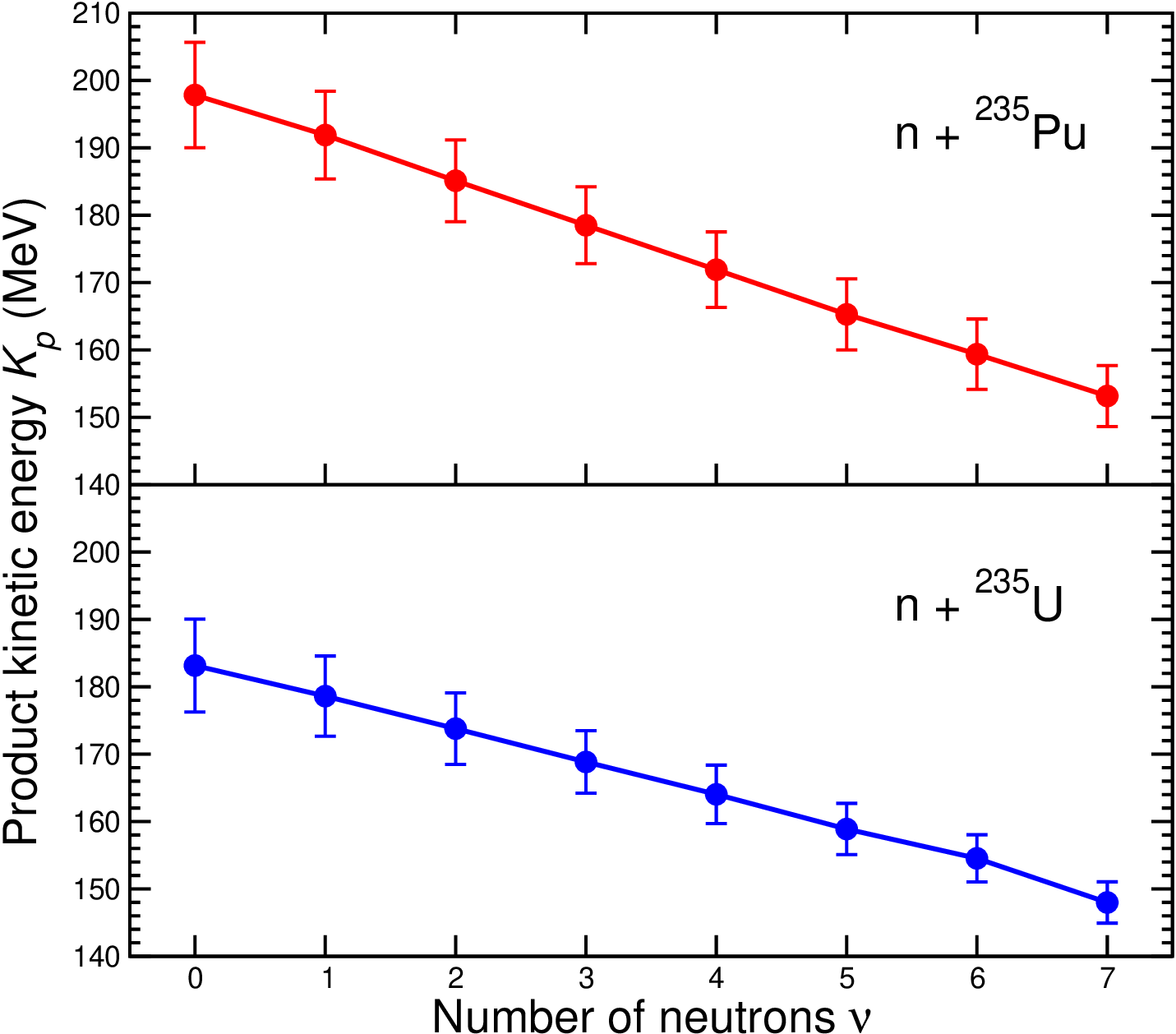}
\caption{
The mean total kinetic of energy the fission products
together with the associated dispersions (bars),
as a function of the neutron multiplicity in the event,
for $0.53\,\MeV\,{\rm n}$ on $^{235}$U ({bottom}) and $^{239}$Pu ({top}).
[Color online.]}\label{f:mKp}
\end{figure}

Finally, Fig.\ \ref{f:mKp} shows how 
the average total fragment kinetic energy of the fission products
and their excitation depend on the number of evaporated neutrons $\nu$.
The decreasing character of the curves is easily understood
since larger neutron multiplicities tend to arise from
higher fragment excitations, 
which occurs in events with lower kinetic energies.

\begin{table}[b]
\begin{tabular}{cc|cccccccc}
\hline
~& $\nu$ & All & ~~1~~ & ~~2~~ & ~~3~~ & ~~4~~ & ~~5~~ & ~~6~~& ~~7~~\\ 
\hline
&	$L$ & 2.30 & 2.38 & 2.30 & 2.19 & 2.02 \\
Pu &	$H$ & 1.64 & 1.70 & 1.64 & 1.58 & 1.50 & 1.34 & 1.17 \\
&	$L$+$H$	& 1.98 & 2.10 & 2.09 & 2.01 & 1.93 & 1.82 & 1.74 & 1.68 \\
\hline
&	$L$ & 2.18 & 2.22 & 2.17 & 2.05 & 1.85 \\
U &	$H$ & 1.50 & 1.56 & 1.46 & 1.39 & 1.24 \\
&	$L$+$H$	& 1.84 & 1.85 & 1.88 & 1.84 & 1.79 & 1.73 & 1.67 & 1.55 \\
\hline
\end{tabular}
\caption{The mean kinetic energy $\overline{\epsilon}_{\rm n}$ (MeV) of the 
neutrons evaporated from  the light fragment ($L$), the heavy fragment ($H$),
or from either one ($L$+$H$),
as a function of the respective multiplicity $\nu_L$, $\nu_H$, or $\nu$,
in fission events induced by thermal neutrons 
on $^{235}$U ({bottom}) and $^{239}$Pu ({top}).
}\label{t:EnLab}
\end{table}

\subsection{Neutron energies}

\begin{figure}[b]	
\includegraphics[angle=0,width=3.1in]{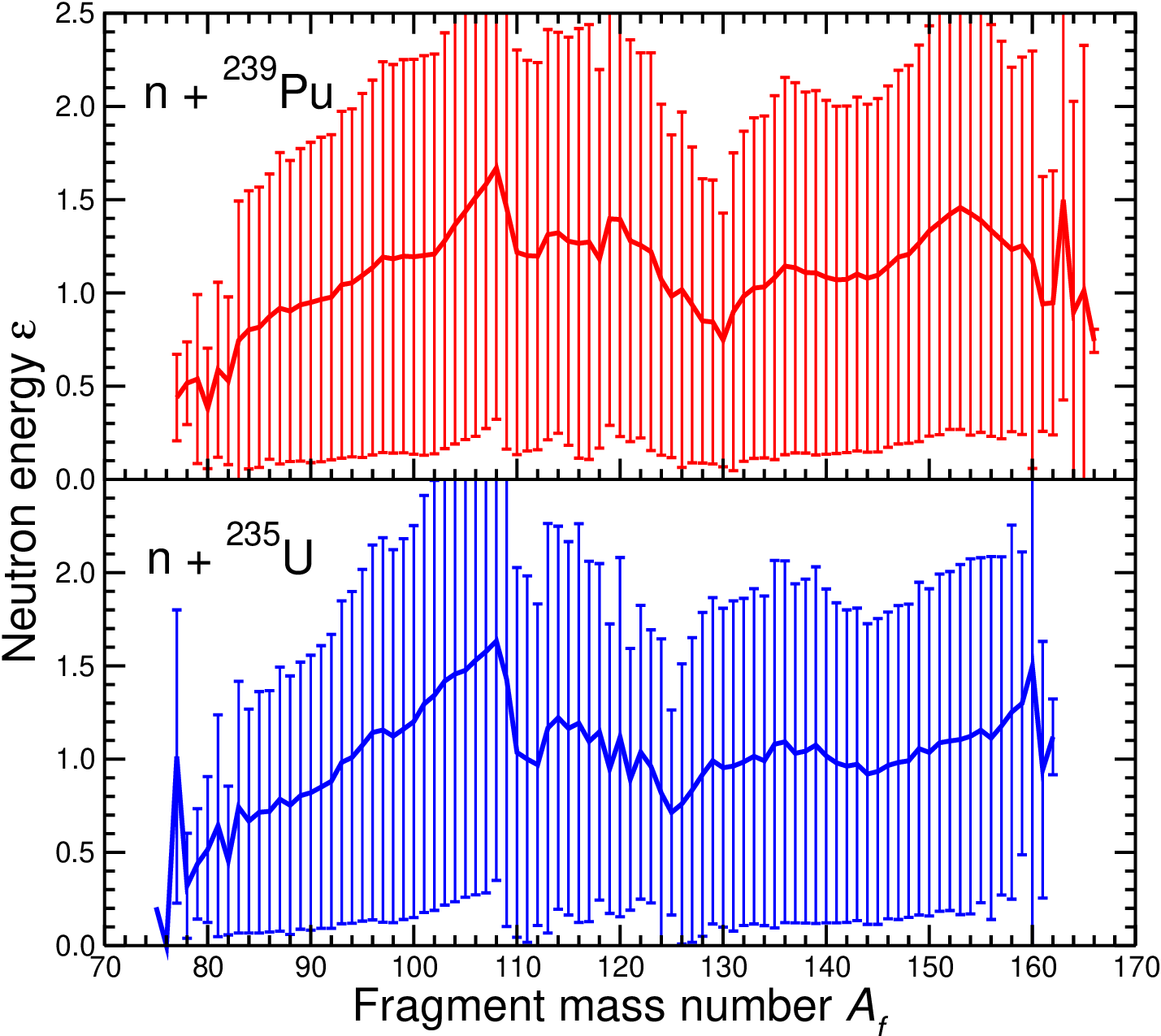}
\caption{
The mean neutron energy $\overline{\epsilon}_{\rm n}$ (curves)
together with its dispersion (bars) as a function of fragment mass $A_f$
for $0.53\,\MeV\,{\rm n}$ on $^{235}$U ({bottom}) and $^{239}$Pu ({top}).
[Color online.]}\label{f:EnA}
\end{figure}

We now turn to the kinetic energies of the evaporated neutrons.
Figure \ref{f:EnA} shows the fragment-mass dependence of 
the mean kinetic energy with respect to the frame of the emitting nucleus
together with the associated dispersion of the kinetic-energy distribution.

The neutron spectra depend somewhat on the number of neutrons emitted.
This is summarized in Table \ref{t:EnLab} which shows the mean
kinetic energy of neutrons emitted from 
the light fragment, the heavy fragment, or from either one,
as a function of the respective neutron multiplicities
$\nu_L$, $\nu_H$, and $\nu=\nu_L+\nu_H$.

\begin{figure}          
\includegraphics[angle=0,width=3.1in]{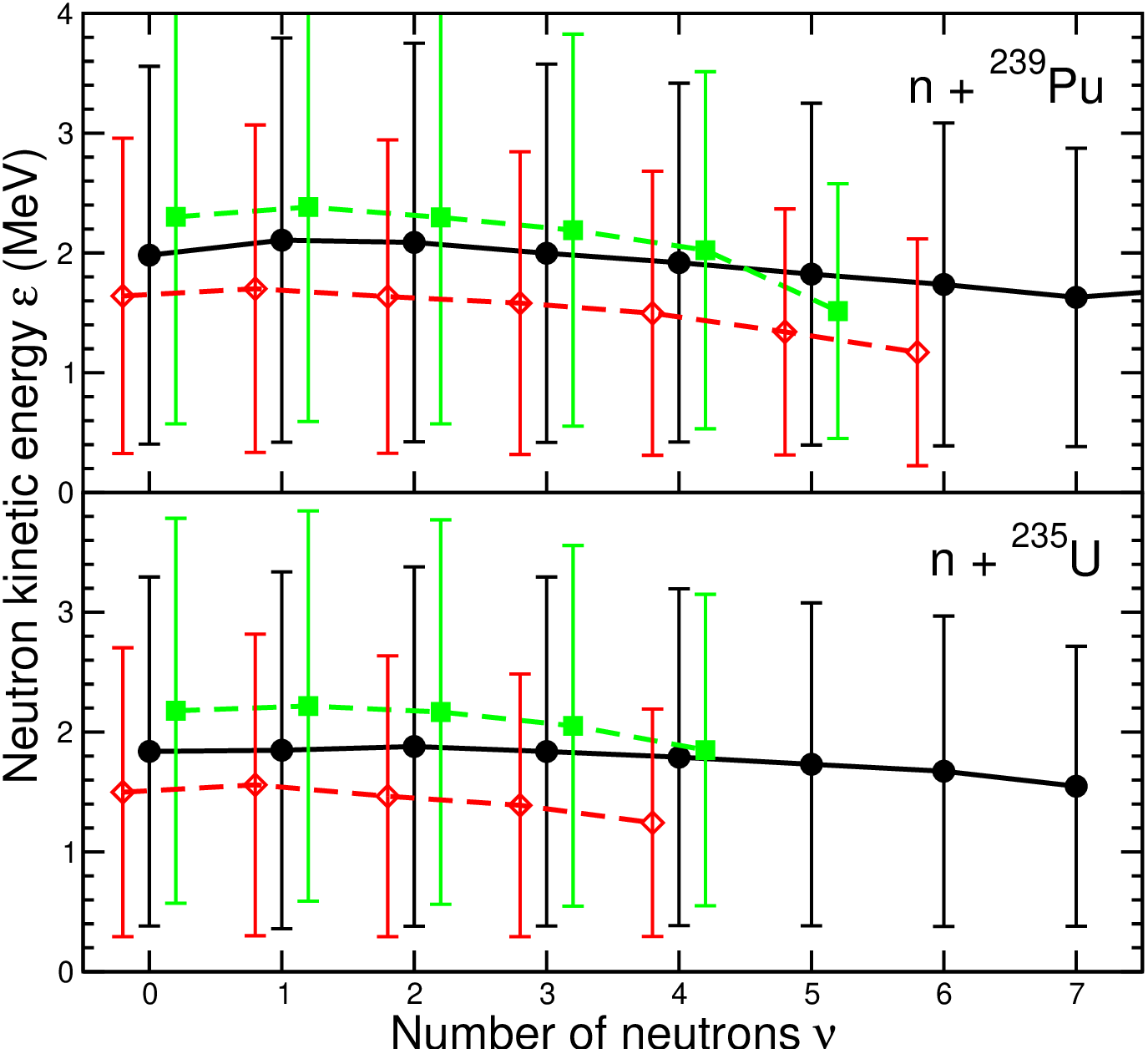}
\caption{
The mean kinetic energy and the associated dispersions  
of all the neutrons emitted in fission events 
with a specified total neutron multiplicity $\nu$, induced by 
thermal neutrons on $^{239}$Pu and $^{235}$U ({\em solid curve, black dots}),
as well as the mean kinetic energy and the associated dispersions  
of all the neutrons emitted from the light fragment
as a function of the corresponding multiplicity $\nu_L$ 
({\em dashed curve, green squares})
and the mean kinetic energy and the associated dispersions  
of all the neutrons emitted from the heavy fragment
as a function of the corresponding multiplicity $\nu_H$ 
({\em dashed curve, red diamonds}).
[Color online.]}\label{f:EnLab}
\end{figure}

The mean energies, as seen in the laboratory, 
as well as the associated dispersions, 
are displayed in Fig.\ \ref{f:EnLab} for the three neutron categories.
In each case,
there is an overall relatively modest decrease of the average neutron energy
(and a corresponding narrowing of the distribution)
as the neutron multiplicity is increased.
This feature would be expected 
since the available energy must be shared among more neutrons.

\begin{figure}          
\includegraphics[angle=0,width=3.1in]{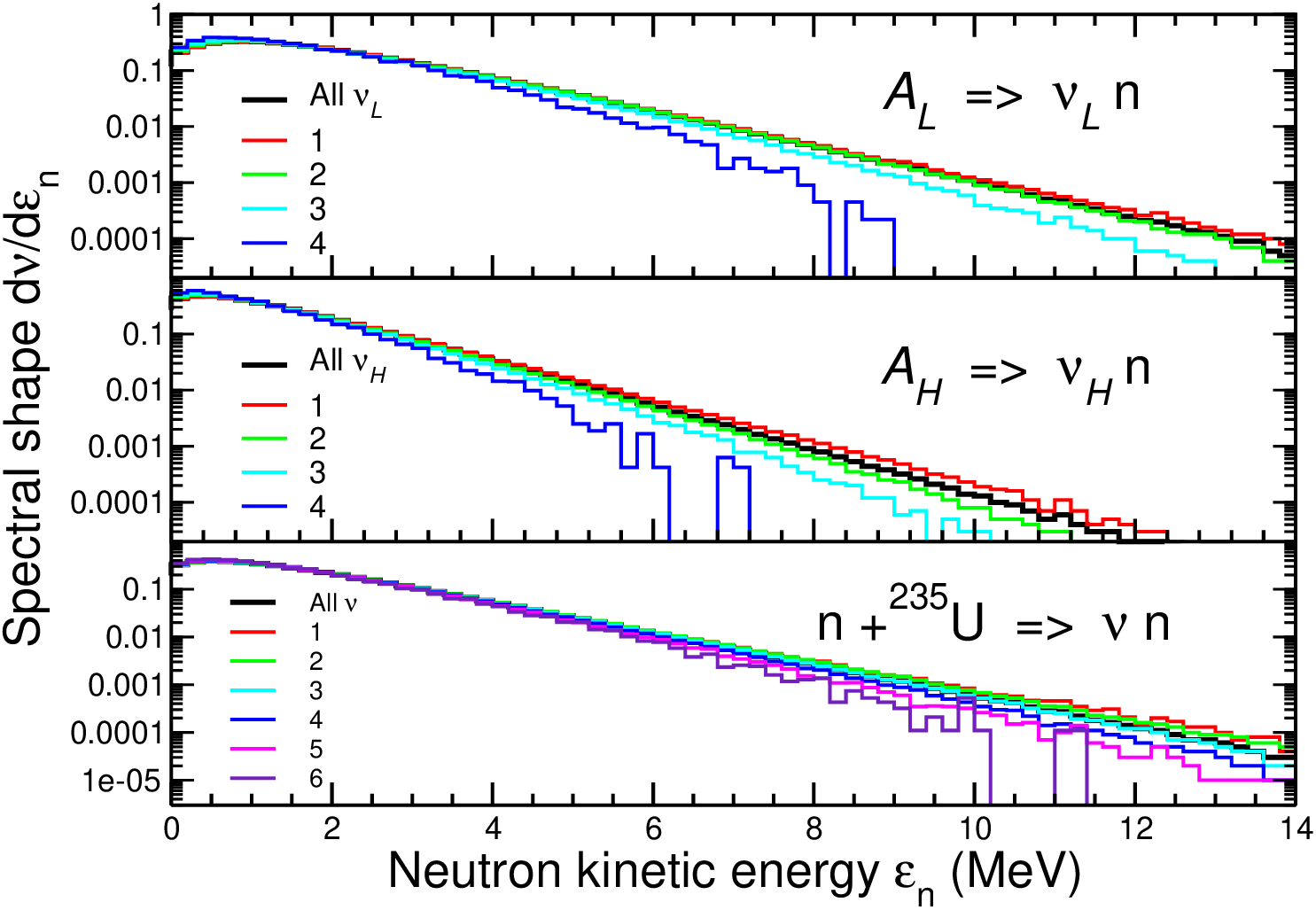}
\caption{
The spectral shape of neutrons evaporated 
from the light ({top}), 
the heavy ({middle}),
or either ({bottom}) fragment
for specified values of the respective multiplicity $\nu_L$, $\nu_H$, or $\nu$,
in fission induced by thermal neutrons on $^{235}$U.
[Color online.]}\label{f:EnLabU}
\end{figure}

\begin{figure}          
\includegraphics[angle=0,width=3.1in]{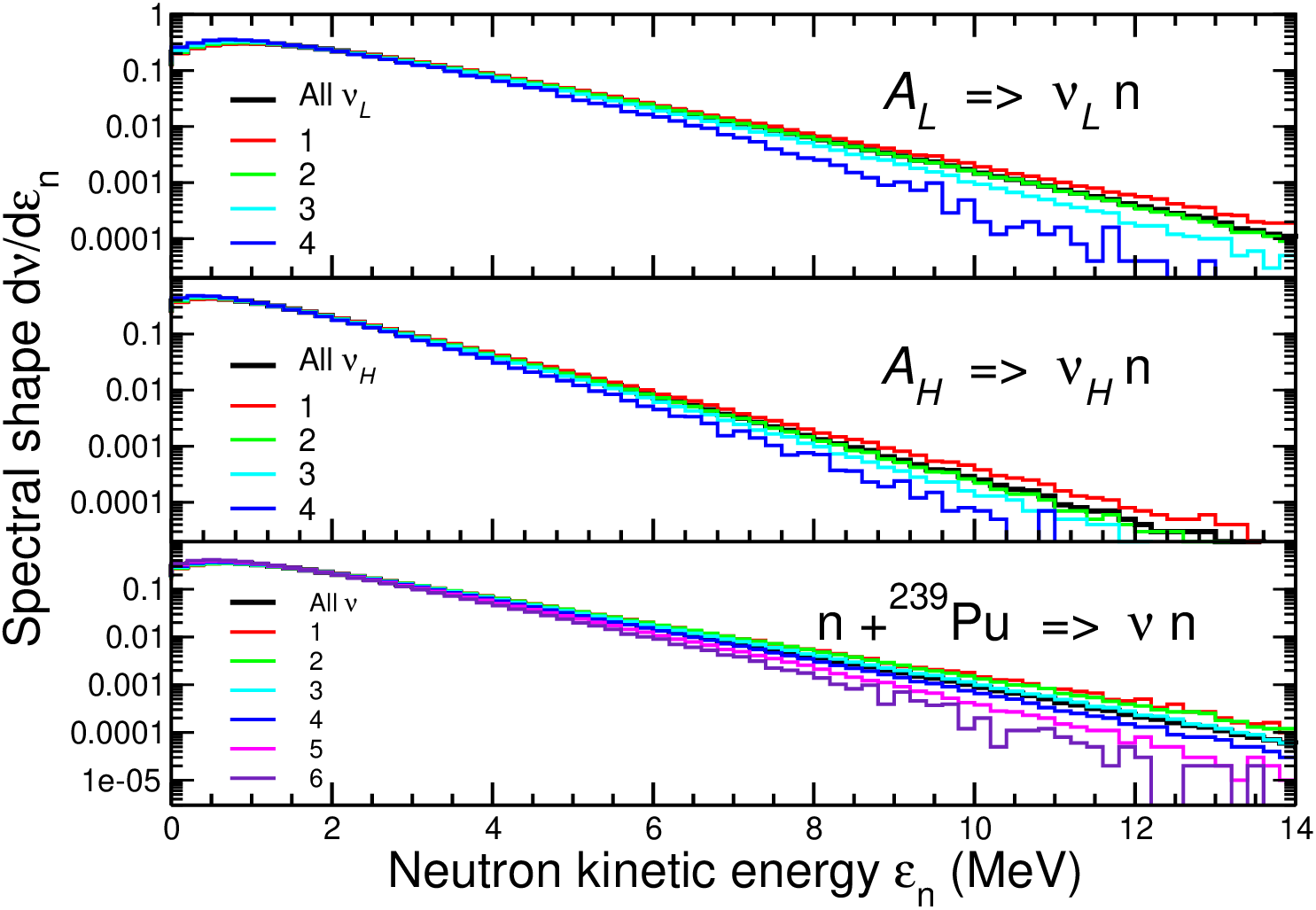}
\caption{
Similar to Fig.\ \ref{f:EnLabU} but for ${\rm n}+^{239}$Pu.
[Color online.]}\label{f:EnLabPu}
\end{figure}

The full multiplicity-gated spectral shapes are shown in 
Figs.\ \ref{f:EnLabU} (for U) and \ref{f:EnLabPu} (for Pu).
It is apparent that the spectra become progressively softer 
at higher multiplicities.
This type of information is not provided by the standard models
and is therefore novel.

\subsection{Neutron-neutron angular correlations}

The event-by-event calculation makes it straightforward
extract the angular correlation between two evaporated neutrons,
an observable that has long been of experimental interest
(see, for example, Refs.\ \cite{DeBenedettiPR74,FranklynPLB78,GagarskiBRAS72}
and references therein)
but which cannot be addressed with the standard models of fission.

\begin{figure}[h]	
\includegraphics[angle=0,width=3.1in]{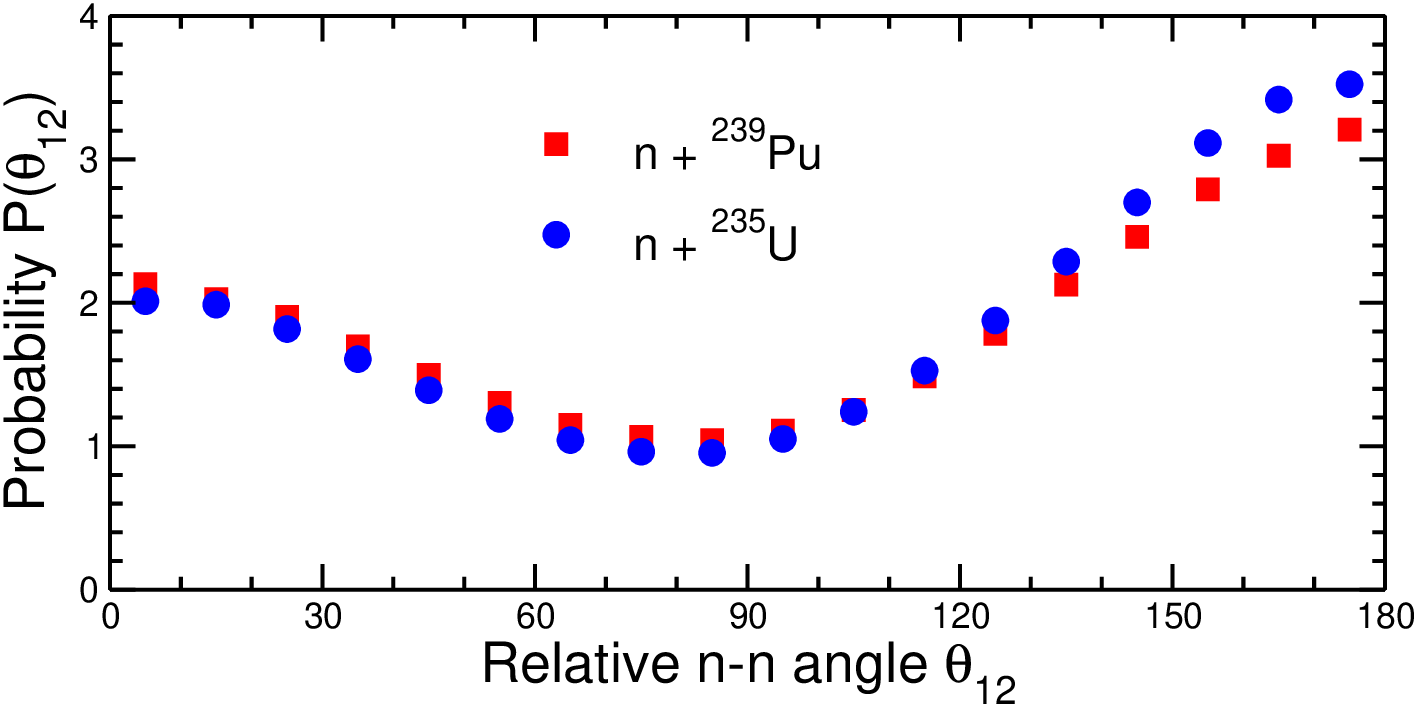}	
\caption{
The angular correlation between two evaporated  neutrons
for $0.53\,\MeV\,{\rm n}$ on $^{235}$U and $^{239}$Pu,
considering only neutrons with a kinetic energy above $1\,\MeV$.
[Color online.]}\label{f:nnCorr}
\end{figure}

Figure \ref{f:nnCorr} shows this quantity for the neutrons resulting from
fission induced by thermal neutrons on $^{235}$U and $^{239}$Pu.
The analysis shown included only neutrons with kinetic energy 
above a threshold of $1\,\MeV$.
The results look qualitatively similar for other threshold energies,
with the angular modulation growing somewhat more pronounced 
as the threshold is raised 
(while the counting statistics is correspondingly reduced).

We see that the neutrons tend to be either forward or backward correlated.
The backward correlation appears to be somewhat favored,
as would be expected from the relatively small but negative value of the
multiplicity correlation coefficient $C_{LH}$ shown in Table \ref{t:nu}.

\subsection{Neutron-photon correlations}

\begin{figure}[b]          
\includegraphics[angle=0,width=3.1in]{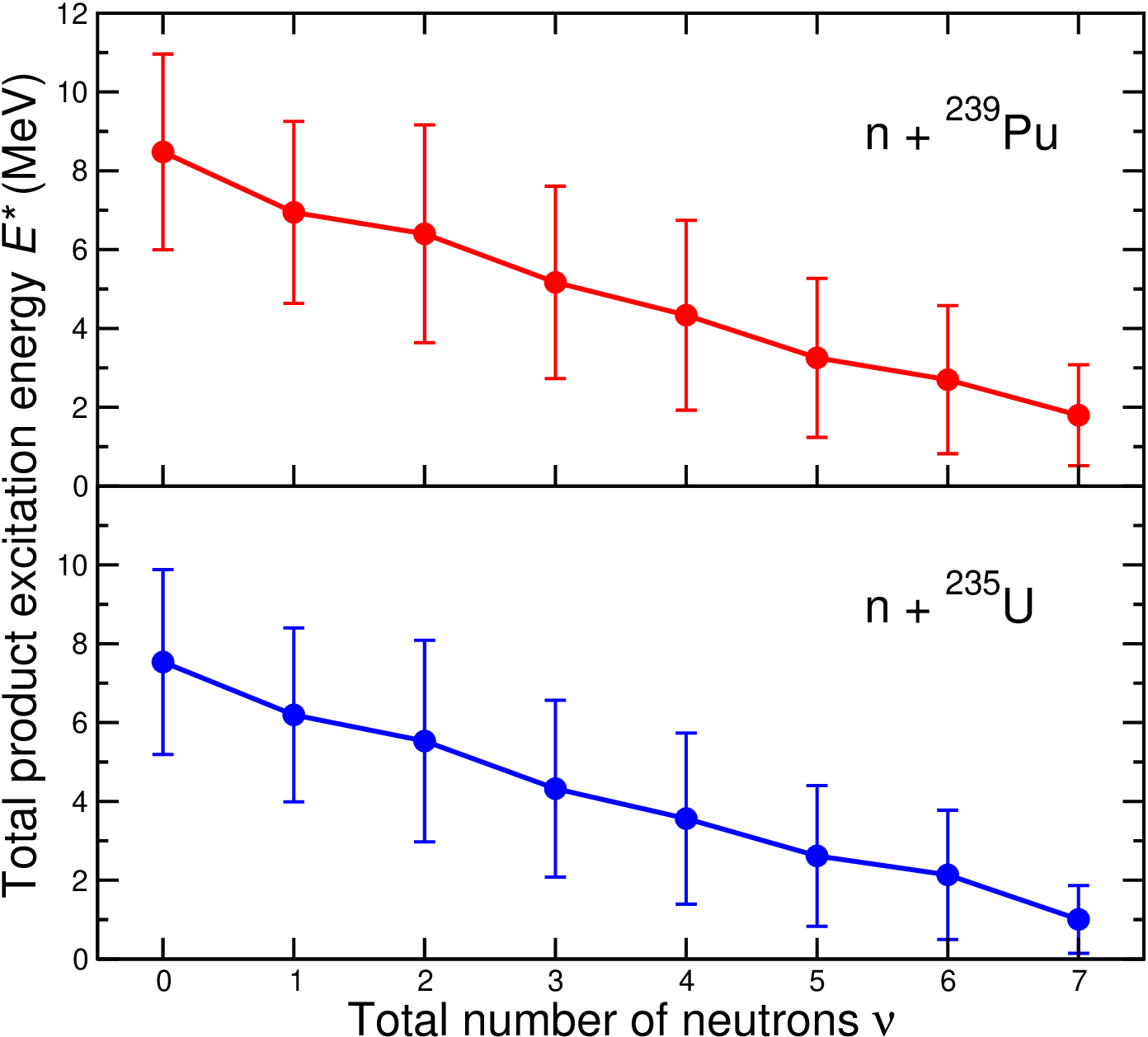}	
\caption{
The mean total excitation of energy the two fission products,
together with the associated dispersions (bars),
as a function of the neutron multiplicity in the event,
for $0.53\,\MeV\,{\rm n}$ on $^{235}$U ({bottom}) 
and $^{239}$Pu ({top}).
[Color online.]}\label{f:mEp}
\end{figure}

The final illustration is relevant for 
the correlation between the neutron and photon multiplicities.
Figure \ref{f:mEp} shows the combined excitation left in the two product nuclei
as a function of the total number of evaporated neutrons.
When more neutrons are emitted
the residual product nuclei are less excited.
This feature appears to be reasonable since a larger-than-average number of
neutrons is likely to have used up a larger-than-average portion of the
total available excitation energy, thus leaving a less-than-average amount
of residual excitation.

Since the average number of photons emitted from a given product
increases monotonically with excitation,
the results in Fig.\ \ref{f:mEp}
provides a qualitative indication of the correlation between
the number of neutrons evaporated and the number of photons emitted
during the further deexcitation of the product nuclei.
Our simulations thus suggest that the two multiplicities are anticorrelated:
the more neutrons the fewer photons.

This qualitative expectation is borne out by Table \ref{t:ng}
which summarizes the result of including the actual photon multiplicity $\mu$
into the analysis.
The covariance between $\nu$ and $\mu$ is indeed negative and
the corresponding correlation coefficient is about minus one third,
suggesting a fairly significant degree of anticorrelation.

\begin{table}
\begin{tabular}{c|cccccc}
\hline
~ 	& ~~~$\overline{\nu}$~~~ & ~~~$\sigma_\nu$~~~
	& ~~~$\overline{\mu}$~~~ & ~~~$\sigma_\mu$~~
	& ~~~$\sigma_{\nu\mu}$~~ & ~$C_{n\gamma}$~~\\
\hline
~n+$^{239}$Pu ~& 2.97 & 1.03 & 5.67 & 2.49 & -0.84 & -0.33\\
~n+$^{235}$U~~  & 2.49 & 0.96 & 5.40 & 2.40 & -0.79 & -0.34\\
\hline
\end{tabular}
\caption{The mean neutron multiplicity $\overline{\nu}=\langle\nu\rangle$ 
and the associated dispersion $\sigma_\nu$,
the mean photon multiplicity $\overline{\mu}=\langle\mu\rangle$ 
and the associated dispersion $\sigma_\mu$,
and the neutron-photon multiplicity covariance
$\sigma_{\nu\mu}\equiv\langle\nu\mu\rangle-\overline{\nu}\overline{\mu}$
together with the corresponding correlation coefficient 
$C_{n\gamma}=\sigma_{\nu\mu}/[\sigma_\nu\sigma_\mu]$,
for photons with transition energies above $200~{\rm keV}$.
}\label{t:ng}
\end{table}

\section{Concluding remarks}
\label{conclude}

Over the last few years,
experimental capabilities have improved dramatically
while the practical applications of fission have broadened significantly.
As a consequence, there has been an growing need for calculations 
of increasingly complex observables that are beyond the scope of the
traditional models employed in the field.

To meet this need, we have developed a new calculational framework, \code,
which can generate large samples of individual fission events.
From those it is then possible to extract 
any specific correlation observable of interest,
without the need for further approximation.
In developing \code, we have sought to make the numerics sufficiently fast
to facilitate use of the code as a practical calculational tool.
(Thus, on a MacBook laptop computer, it takes about 12~seconds
to generate one million events.)

Our early emphasis has been on creating a working code
that can produce samples of reasonably realistic fission events
and form a convenient basis for gradual refinements.
(Its simple modular structure should facilitate such further developments.)
Consequently, the present version is still rather rough and cannot compete
for quantative accuracy with established models,
without suitable {\em ad hoc} parameter adjustments (see Ref.\ \cite{VPRY}).
Even so, the model has already proven to be capable of making interesting 
predictions for correlations of interest in variety of contexts
and we foresee an increased number of applications.

\section*{Acknowledgements}
We wish to acknowledge helpful discussions with 
D.A.\ Brown, D.\ Gogny, E.\ Ormand, P.\ M{\"o}ller,
E.B.\ Norman, J.\ Pruet, W.J.\ Swiatecki, P.\ Talou, and W.\ Younes.
This work was supported by the Director, Office of Energy Research,
Office of High Energy and Nuclear Physics,
Nuclear Physics Division of the U.S. Department of Energy
under Contracts No.\ 
DE-AC02-05CH11231 (JR) and
DE-AC52-07NA27344 (RV) 
and by the National Science Foundation, Grant NSF PHY-0555660 (RV).

\appendix
\section{Liquid-drop model}
\label{LD}

For simplicity, we use here a the liquid-drop model \cite{MyersNPA81}
for the macroscopic part of the nuclear binding energy.
Accordingly, the surface and Coulomb energy of a spherical nucleus 
are given by
\beqar
E_S^0(^AZ) &=& a_2 A^{2/3}[1-\kappa\left({N-Z\over2A}\right)^2]\ ,\\
E_C^0(^AZ)  &=& c_3{Z^2\over A^{1/3}}\ ,
\eeqar
with the Lysekil parameter values:
$a_2=17.9439\,\MeV$,
$\kappa=1.7826$,
$c_3=\mbox{$3\over5$}e^2/r_0=0.7053\,\MeV$  \cite{Lysekil}.

The distortion energies of the prefragments at scission
are based on the shape dependence of the surface and Coulomb energies
of macroscopic prolate nuclei \cite{WDM:Book}
\beqar
E_S(\eps) &=&  E_C^0 B_S(\eps)\ \approx\ 
E_S^0[1+\mbox{$8\over45$}\eps^2]\ ,\\
E_C(\eps) &=&  E_C^0 B_S(\eps)\ \approx\ 
E_C^0[1-\mbox{$4\over45$}\eps^2]\ .
\eeqar

\section{Level densities}
\label{aA}

The relationship between the nuclear excitation energy $E^*$
and the nuclear temperature $T$ is generally somewhat complicated.
For the time being, we simply use the familiar approximation 
$T=\sqrt{\eps^*/a}$, where $\eps^*=E^*-\delta E_{\rm def}$ 
is the statistical part of the excitation energy (the ``heat'').
For a given nucleus $(Z_i,A_i)$ the level-density parameter $a_i$ 
is taken from Ref.\ \cite{KouraJNST43},
\beq
a_i(E^*)\ =\ {A_i\over e_0}\left[1+{\delta W_i\over U_i}\,
(1-\rme^{-\gamma U_i})\right] ,\ U_i\equiv E^*-\Delta_i\ ,
\eeq
with $e_0=7.25\,\MeV$ and $\gamma=0.05$.
Here $\tilde{a}_i=A_i/e_0$ is the asymptotic level-density parameter
whose parameter $e_0$ depends slightly 
on the specific value used for the damping coefficient $\gamma$.
The shell correction energies $\{\delta W_i\}$ 
and the pairing energies $\{\Delta_i\}$
are those calculated by Koura \etal\ \cite{KouraNPA674}
for nuclei with $20\leq Z_i\leq92$.
We note that $a_i(E^*\approx\Delta)\approx
\tilde{a}_i\{1+\delta_i\gamma[1-\half(E^*-\Delta_i)]\}$ is regular.
Furthermore, $a_i(E^*=0)\approx\tilde{a}_i[1+\gamma\delta W_i]$
when $\gamma U_i\ll1$, which is most often the case.
Finally, as $E^*$ is increased we have
$a_i(E^*)\to\tilde{a}_i[1+\delta W_i/E^*]\to\tilde{a}_i\equiv A_i/e_0$.

\section{Spectral sampling}
\label{eps}

It is possible to devise a fast algorithm for sampling 
the spectral distribution (\ref{dNdeps}) for the evaporated neutron,
$dN/d\epsilon\propto\epsilon\,{\rm e}^{-\epsilon/T}$.
It is based on the observation that the function $x{\rm e}^{-x}$ 
is a (normalized) Poisson distribution of order 2.
Hence it can be expressed as the convolution of two (normalized) exponentials
(each of which is a Poisson distribution of order 1), $P_2=P_1*P_1$
with $P_1(x)\equiv{\rm e}^{-x}$,
\beq
x{\rm e}^{-x} =
\int_0^\infty\!dx_1 \int_0^\infty\!dx_2\, 
	\delta(x_1+x_2-x)\, {\rm e}^{-x_1}{\rm e}^{-x_2}\ .
\eeq
This is a special case of the general feature of Poisson distributions,
$P_{n+m}=P_n*P_m$.

We may therefore obtain a sampled value of the kinetic energy $\epsilon$
as the sum of two energies, $\epsilon_1$ and $\epsilon_2$, 
that have each been sampled from a usual exponential distribution 
$\propto{\rm e}^{-\epsilon_i/T}$.
Since the sampling from an exponential distribution $p(x)={\rm e}^{-x}$
 is readily accomplished by sampling a random number
$\eta$ that is uniformly distributed on the interval $(0,1]$ 
and then taking the negative of its logarithm, $x=-\ln\eta$,
the relative neutron kinetic energy is
\beq
\epsilon_{\rm n} = \epsilon_1+\epsilon_2
= -[\ln\eta_1+\ln\eta_2]T_f^{\rm max}\ ,\ \eta_i\in(0,1]\ ,
\eeq
where the two numbers $\eta_i$ have been sampled from $(0,1]$.
Since both mean values and variances are additive under convolution
and each exponential distribution yields
$\langle\epsilon_i\rangle=T$ and $\sigma^2({\epsilon_i})=T^2$,
the resulting relative kinetic energy $\epsilon_{\rm n}$ 
has the mean value $\langle\epsilon_{\rm n}\rangle=2T_f^{\rm max}$ 
and the variance $\sigma^2({\epsilon_{\rm n}})=2(T_f^{\rm max})^2$,
for a fixed value of $T_f^{\rm max}$.

The energy spectrum of the post-evaporation photons
can be sampled rapidly in an analogous manner,
since the corresponding spectral shape,
$dN/d\epsilon\propto\epsilon^2\,{\rm e}^{-\epsilon/T_f}$,
is (proportional to) a Poisson distribution of order 3,
So
\beq
\eps_\gamma = -[\ln\eta_1+\ln\eta_2+\ln\eta_3]T_f\ ,\ \eta_i\in(0,1]\ .
\eeq
It also follows that the mean value is
$\langle\epsilon_\gamma\rangle=3T_f$ 
and the variance is $\sigma^2({\epsilon_\gamma})=3T_f^2$,
for a fixed value of $T_f$.

\section{Lorentz boost}
\label{boost}

We describe here the Lorentz boost required to express the motion
of an ejectile and the corresponding daughter nucleus
in the adopted reference frame.

The boost velocity is that of the mother nucleus,
$\bold{V}_i=\bold{P}_i/E_i$, where $\bold{P}_i$ is the momentum of the
mother nucleus and $E_i$ is its total energy, $E_i^2=(M_i^*)^2+P_i^2$.
To perform the Lorentz boost,
we first note that the component of the ejectile momentum parallel
to the boost velocity is $p_{\rm n}^\parallel=\bfp_{\rm n}\cdot\bold{\hat{v}}$
where $\bold{\hat{v}}\equiv\bold{V}_i/V_i$ 
is the unit vector in the direction of $\bold{V}_i$.
The component transverse to $\bold{V}_i$ is then
$\bfp_{\rm n}^\perp=\bfp_{\rm n}-p_{\rm n}^\parallel\bold{\hat{v}}$ and
this component is unaffected by the boost, 
$\tilde{\bfp}_{\rm n}^\perp=\bfp_{\rm n}^\perp$.
The parallel component of the ejectile momentum and its energy 
transform as follows,
\beq
\tilde{p}_{\rm n}^\parallel = \gamma(p_{\rm n}^\parallel+E_{\rm n}V_i)\ ,\
\tilde{E}_{\rm n} = \gamma(E_{\rm n}+\bfp_{\rm n}\cdot\bold{V}_i)\ ,
\eeq
where $\gamma^2=1/(1-V_i^2)$ and $E_{\rm n}^2=m_{\rm n}^2+p^2$.
Thus the boosted ejectile momentum is
\beq
\tilde{\bfp}_{\rm n}\ =\ 
[\gamma E_{\rm n}+{\gamma-1\over V_i^2}\,\bfp_{\rm n}\cdot\bold{V}_i]
\bold{V}_i\ + \bfp_{\rm n}\ ,
\eeq
while the boosted value of the recoil momentum is
obtained by reversing the direction of $\bfp$,
\beq
\tilde{\bold{P}}_f\ =\ 
[\gamma E_f-{\gamma-1\over V_i^2}\,\bfp_{\rm n}\cdot\bold{V}_i]\bold{V}_i\ 
- \bfp_{\rm n}\ ,
\eeq
with $E_f$ being the total energy of the daughter, $E_f^2=(M_f^*)^2+p^2$.


~\hfill LLNL-JRNL-413625

			\end{document}